\documentclass[5p,number]{elsarticle}
\biboptions{sort&compress}

\usepackage{amsmath}
\usepackage{upgreek}

\journal{JCG}

\begin{document}

\begin{frontmatter}

\title{Growth of different faces in a body centered cubic lattice: a case of the phase-field-crystal modeling}


\author[add2,add4,add1]{V.~Ankudinov\corref{cor1}}
\ead{vladimir@ankudinov.org}

\author[add1,add3]{P.\,K.~Galenko}

\address[add2]{Vereshchagin Institute of High Pressure Physics, Russian Academy of Sciences, 108840~Moscow (Troitsk), Russia}
\address[add4]{Udmurt Federal Research Center, Ural Branch of the Russian Academy of Science, 426000 Izhevsk, Russia}
\address[add1]{Udmurt State University, Institute of Mathematics, Informatics and Physics, Condensed Matter Physics Lab, 426034 Izhevsk, Russia}
\address[add3]{Friedrich-Schiller-Universität, Otto-Schott-Institut für Materialforschung, Lehrstuhl Metallische Werkstoffe, Löbdergraben 32, 07743 Jena, Germany}


\cortext[cor1]{Corresponding author}

\begin{abstract}
Interface energy and kinetic coefficient of crystal growth strongly depend on the face of the crystalline lattice. To investigate the kinetic anisotropy and velocity of different crystallographic faces we use the hyperbolic (modified) phase field crystal model which takes into account relaxation of atomic density (as a slow thermodynamic variable) and atomic flux (as a fast thermodynamic variable). Such model covers slow and rapid regimes of interfaces propagation at small and large driving forces during solidification. In example of BCC crystal lattice invading the homogeneous liquid, dynamical regimes of crystalline front propagating along the selected crystallographic directions are studied. The obtained velocity and the velocity sequences for different faces are compared with known results.
\end{abstract}

\end{frontmatter}

\newpage


\section{Introduction}

The phase-field crystal model (PFC) was formulated \cite{bib:pe,elder12} to describe continuous transitions from the homogeneous to the periodic state (similarly to the Landau–Brazovskii transition \cite{Landau1996,brazovskii75,bib:L}) and between different periodic states evolving over diffusion times. The model is based on the description of the free energy, which is a functional of the atomic density field periodic in the solid phase and homogeneous in the liquid disordered state. The form of the free energy close to the Swift-Hohenberg type \cite{ge-swift} and allows to account structural phase transitions of the first and second order.

Recent advances in PFC-modeling of the different aspects of crystallization allow one to model many scenarios such as dynamics of freezing of colloids and polymers, epitaxial growth, ordering of the structures on nano-scales~\cite{bib:adv2012} and  rapid crystallization~\cite{GDL}. Results of PFC simulations could provide interface energies, pattern selection under non-equilibrium conditions and velocities of moving phase boundaries~\cite{granasy}. As a simplification of density functional technique (DFT) for freezing~\cite{rama79}, a PFC model  utilizes several approximations~\cite{elder12,bib:berry2008} which makes it relevant for modelling of  nucleation from undercooled liquids~\cite{Guerdane2018}, dendritic crystal growth~\cite{Toth2010, Tang2011}, heteroepitaxy and multi-grain growth in presence of hydrodynamical flows~\cite{Podmaniczky2017}.

In the present work, we numerically investigate rapid growth of different faces in body centered cubic crystal lattice (BCC-lattice). With this aim we use the hyperbolic (modified) PFC-model which takes into account relaxation of the phase field and relaxation of the flux of atomic density \cite{book,GDL}. In particular, an influence of the local relaxation time and the effect of atomic reticular density of different crystal faces on the interface velocity are studied.

\section{PFC model}
The hyperbolic (modified) PFC-model (MPFC) for fast transitions includes the inertial term for the atomic density \cite{book,GDL} as the result of accounting for the flux $\vec{J}(\vec{r},t)$ and atomic density  field $n(r,t)=\rho(r,t)/\rho_0-1$ in a form of independent thermodynamic variables, where $\rho(r,t)$ is a density field and $\rho_0 $ is a reference density. Therefore, in MPFC, the free energy is a functional of $n$ and $\vec{J}(\vec{r},t)$ formulated as \cite{GDL,ge,book}:
\begin{align}
F(n,\vec{J})=\int \left[\frac{n}{2} \left( - \varepsilon +  \mathcal{D}_i\right) n - \frac{a}{3}n^3 + \frac{v}{4} n^4 \right] d\vec{r}  + \nonumber \\
+ \frac{\tau}{2} \int (\vec{J}\cdot \vec{J} \,) d \vec{r},
\label{free}
\end{align}
where $a$ and $v$ are phenomenological constants which control phase transition, $\tau$ is the relaxation time of the atomic density flux $\vec{J}(\vec{r},t)$, $\varepsilon$ is the driving force (quench depth) and the differential operator $\mathcal{D}_i$  \emph{per se} elastic module  \cite{zaeem16,ankudinov17iop-2,ankudinov16,bib:pe} describes  crystallographic symmetries in one-mode ($i=1$) and two-mode ($i=2$) approximations:
\begin{equation}
\label{op}
 \mathcal{D}_i =
 \begin{cases}  r_0+(q_0^2+\nabla^2)^2, & \quad   i=1,\\
 [r_0+(q_0^2+\nabla^2)^2][r_1+(q_1^2+\nabla^2)^2], & \quad i=2. \end{cases}
\end{equation}
The coefficients $r_0$ and $r_1$ are responsible for the relative stability of crystalline structures, and $q_0$ and $q_1$ are the  modules  of first two  sublattice wave vectors \cite{asadi15,ankudinov17iop-2,Tang2011}. In current work we utilize one-mode case $i=1$ with $r_0=0$, $q_0=1$ which corresponds to the stable BCC structure \cite{asadi15} in one-mode case.

The dynamical MPFC-equation for atomic density field is \cite{GDL,ge,book,provatas,bueno16}: 
\begin{equation}
\tau \frac{\partial^2 n}{\partial t^2} +\frac{\partial n}{\partial t}=\nabla^2 \mu, \quad \quad \mu = \frac{\delta F}{\delta n},
\label{dyn}
\end{equation}
where $ \mu (n) $ is a chemical potential defined by the functional derivative of the free energy Eq.~(\ref{free}),
\begin{equation}
\mu=(1-\varepsilon) n - a n^2 + v n^3 + 2 \nabla^2 n + \nabla^4 n.
\label{mumumu}
\end{equation}
If the relaxation of the flux proceed infinitely fast, $\tau\rightarrow 0$, Eq.~(\ref{dyn}) transforms to the original parabolic PFC-equation  \cite{Elder2007,ggke-voigt,bib:epbsg07,elder12,bib:berry2008,granasy,bib:adv2012}. The inertia term  $\tau\partial^2 n / \partial t^2$  makes the hyperbolic dynamics possible to capture the fast propagation front over the times $\tau$ for flux relaxation~\cite{book,GDL,ge} that is important to predict non-equilibrium effects, for instance, in rapid solidification~\cite{gj19}. Alternatively, Eq.~(\ref{dyn}) was proposed in Ref.~\cite{provatas} to incorporate both mass diffusion and fast elastic relaxation. Proposed hyperbolic term (see Eq.~\ref{dyn}) could be obtained with the past history approach, using the memory kernel to account the history of the force $\nabla \mu$~\cite{GDL,book,Conti2016}.
The equilibrium properties of solid-liquid interfaces such as interfacial energies and its anisotropy were studied previously~\cite{Wu2006,wu07} for the case of small $\varepsilon$'s.

Using Eqs.~(\ref{dyn})-(\ref{mumumu}), we simulate the growth of different faces in BCC-lattice at high\footnote{Under high driving forces we assume the range at which equilibrium values of lattice parameter does not allow one to predict the interface velocity consistently with experimental data or numerical modelling. See Fig.~4 in \cite{gse}, where this regime was achieved for dimensionless driving force $\varepsilon>0.2$.} driving forces $\varepsilon>0$ with non-zero values of the relaxation time $\tau$ at $a=0$. More specifically, we analyze the crystallization fronts moving from metastable liquid for seeds with different initial orientations of the BCC-lattice.

\subsection{Numerical solution of MPFC equation}
Introducing the new variable $P$ one can split Eqs.~(\ref{dyn})-(\ref{mumumu}) effectively reducing the spatial derivative order~\cite{ge}:
\begin{align}
\left\{
\begin{array}{ll}
\tau \dfrac{\partial^2 n}{\partial t^2}  +\dfrac{\partial n}{\partial t}=\nabla^2 \mu, \\
 \mu  =  (1-\varepsilon) n - a n^2 + v n^3 +  \nabla^2 (2 n +  P)  \\
 P =   \nabla^2 n
\end{array}
\right.
\label{num}
\end{align}
The numerical simulation of crystallization in computational domain of slab configuration was initiated by introducing the periodic crystalline nucleus oriented along the elongated side of the domain (which was merging with the direction of growth, see Fig.~\ref{defect}). The amplitude of density field $n$ for initial seed was set as a constant $\eta=0.5$. The seed's equilibrium lattice parameter $2\pi/q$  was found from the free energy minimization~\cite{ankudinov17iop-2,ankudinov16}. The equilibrium modulus of wave vector was set $q=\sqrt{2}/2$.  The initial average density $n_0$ was found for each initial driving force $\varepsilon$ along the melting line on structure diagram.  Regions of phase existence were found also using the free energy minimization~\cite{ankudinov17iop-2,ankudinov16}. The computational domain consists of $27\times27\times300$ dimensionless units with up to $230$ grid points alongside; the maximum tetrahedral mesh element size was set $\ell=1.4$. Additional tests was performed with domain size $45\times45\times400$, $\ell=1.5$. The lateral size of the domain matched the equilibrium lattice parameter to reduce the possible strain effects. We compared the obtained front positions on this different size domains and found no influence of box size on the results. Such domain size was enough to reach the constant interface velocity with the formation of periodic crystalline lattice. The periodic boundary conditions were defined at the elongated sides, with the isolation on the other surfaces of the slab. We got no huge difference in the value of front velocity for periodic and non-periodic boundary conditions except the stability and periodicity of formed crystal.   The position of the solid-liquid interface in time was obtained by the determination of position of the minimum of density field time derivative $\partial n/\partial t$ which corresponds to the highest gradients of the chemical potential $\mu$. This approach allow one to naturally locate the interfaces  in PFC-models and to evaluate the position of physically justified boundaries.

For the second time derivative, we utilize a solver with the backward differentiation formula having accuracy up to $5^{th}$ order of magnitude. The system of equations~(\ref{num}) has been solved numerically in three-dimensions using a direct scheme of the finite element method with the Lagrange-$C^2$ elements utilizing the COMSOL Software~\cite{comsol2} on two-processor Xeon-based computer.

\section{Results and discussion}
\subsection{Front velocities}
\begin{figure}
  \includegraphics[width=\columnwidth]{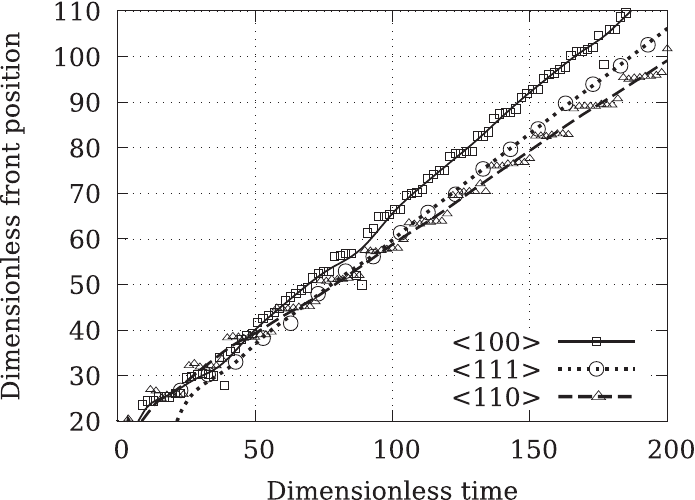}
  \caption{\label{frontPos} BCC solid-liquid interface position vs dimensionless time for different orientation of initial seeds, $\varepsilon=0.4$, $n_0=-0.38$, $\tau=0$.} \end{figure}

We studied the free growth of $\langle 100 \rangle$, $\langle 110 \rangle$ and $\langle 111 \rangle$ faces of BCC-lattice as the planar front observed in all our simulations. The average dimensionless velocity $V$ of the faces was obtained from the front positions in time. Its value was defined as $V=V_f/V_0$ with the front velocity $V_f=(Z^\mathbf{r}-Z^\mathbf{r}_0)/(t'-t'_0)$~[m/s]. The velocity scale $V_0$ was obtained from the scales of front position $Z^\mathbf{r}$~[m] and time $t'$~[s] as \cite{Tang2011,wu07}:
\begin{equation}
V_0=  \lambda k_m^5 M_\rho, \quad\quad \lambda=\frac{k_B T \Gamma}{\rho_0^\ast k_m^4},
\end{equation}
where $T$ is the temperature, $k_B$ is the Boltzmann constant, $k_m$ is the position of the first maximum of correlation functional $C(k)$ in the reciprocal space, $\rho_0^\ast $ is the one-particle number density of the reference liquid state, $M_\rho$ is the mobility, $\Gamma$ is the model coefficient determined by the second derivative of correlation function and the position of its first peak. (see \cite{jaatinen09,Tang2011,wu07} for details).  As the result, the slope of the curve "front position -- time" defines the value of the dimensionless velocity $V$.

Figure~\ref{frontPos} clearly shows that the slope of all of three faces after a short period of irregular motion in the beginning becomes constant $V=const$. The averaged interface positions indicated on Fig.~\ref{frontPos} was found by the  B\'ezier smoothing procedure. The non-monotonic point's positions indicate faceted morphology of stepwise growth previously observed by \cite{granasy}. The strongest faceting found for $\langle 110 \rangle$, and less one for $\langle 100 \rangle$. To quantify $V$ we fit the set of the front positions $Z_i$ at $t_i$ to the function $Z=z_0+V t$, where $z_0$ is the initial position. Selecting the moderate front positions $100<Z_i<250$ and correspondent times $t_i$ the data free of the boundary effects have been used.
The uncertainty in the front velocity $\Delta V \simeq \Delta Z / t $. Due to the constant maximum error in front position determination $\Delta Z=8.9$ equal to the lattice parameter in ${\langle 100 \rangle}$ direction for fixed domain size we get the systematic error $\Delta V= 4.2\%$. 
 For the front positions presented on Fig.~\ref{frontPos} we obtained the following values of averaged velocities: $V_{\langle 100 \rangle}=0.519$, $V_{\langle 111 \rangle}=0.4623$, $V_{\langle 110 \rangle}=0.4027$ with the appeared anisotropies $V_{\langle 100 \rangle}/V_{\langle 110 \rangle}\approx1.29$ which are qualitatively agree with the results of work~\cite{Tang2011}.

Gr{\'{a}}n{\'{a}}sy et al.~\cite{granasy,Toth2010} showed the decreasing velocity in time $V\sim t^{-1/2}$, confirming the diffusion-controlled regime of solidification with the existence of different sequence of growth velocities obtained as
$V_{\langle 111 \rangle}>V_{\langle 100 \rangle}>V_{\langle 110 \rangle}$.
In their work the position of front is proportional to square root of time in the beginning of crystallization front movement, which is well observed in the case of small driving forces used in \cite{granasy,Toth2010}.
There is an analysis of front dynamics (see work \cite{Galenko2000a} for a pure element system solidification and the work \cite{Galenko2000} for a binary alloy system) which shows that at a small driving force, the growth velocity will decay in time and the position of the front will be a function of square root of time. However, for the larger driving forces the front velocity should attained the constant value behind some critical driving force $\varepsilon$. A search for this critical value in the PFC is a perspective topic of future study.
The accounting of the cubic term ($a\neq 0$ in \cite{granasy,Toth2010}) in free energy Eq.~(\ref{free}) could also lead to the retention of the front movement to the nonlinear behaviour which may also influence on the comparative faces velocities. In our simulations, the linear dependence of front positions in time has been found with the velocity sequence different from the work of Gr{\'{a}}n{\'{a}}sy et al.~\cite{granasy,Toth2010}: we have obtained for BCC crystals that the most fast growth occurs for the faces with the lowest reticular atomic density in consistency with the crystallographic law of Bravais (see Section~\ref{retdens}).

\begin{figure}
  \includegraphics[width=\columnwidth]{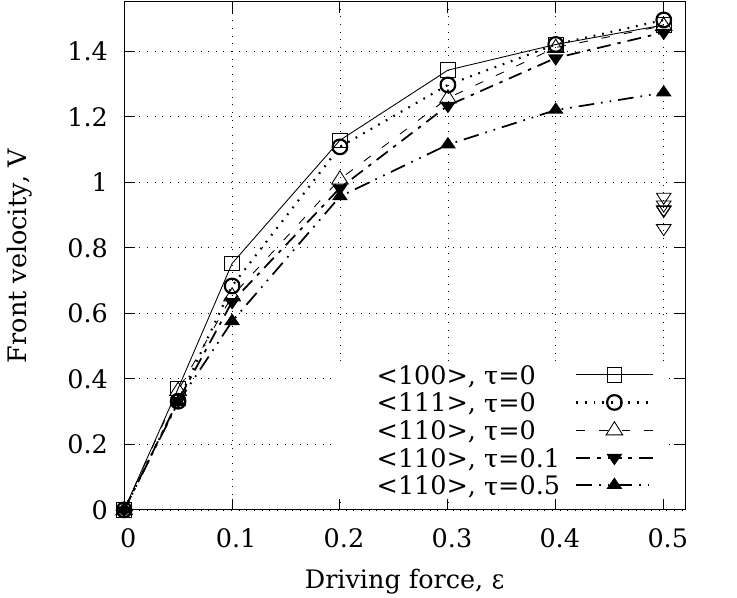}
  \caption{\label{vel} BCC front velocities $V$ as a function of driving force $\varepsilon$ for different crystallographic faces. Represented lines and points correspond to initial densities $n_0=-0.108, -0.13, -0.19, -0.26, -0.31, -0.36$ for driving forces  $\varepsilon=0.05, 0.1, 0.2, 0.3, 0.4, 0.5$ respectively.  Solid triangles correspond to the propagation of $\langle 110 \rangle$-front for $\tau=0.1$ and $\tau=0.5$. The additional group of triangles $\bigtriangledown$ from top to bottom corresponds to the different values of relaxation time $\tau=0.05, 0.1, 0.2, 0.3, 0.5$ for the driving force $\varepsilon=0.5$ and averaged density $n_0=-0.38$. The maximum systematic error is  $\Delta V=4.2\%$.}
  \end{figure}

The non-linear dependence of $V(\varepsilon)$ presented in Fig.~\ref{vel} demonstrates effect of relaxation time on the velocity of different crystallographic faces in BCC-lattice.
 During the solidification of undercooled liquid we observed the spontaneous recrystallization and growth of the rods-phase simultaneously with the BCC-crystal for the large initial densities $n_0>-0.19$ particularly for the $\langle 111 \rangle$ and $\langle 110 \rangle$  faces. One can observe that stable BCC-structure can growth in the nonequilibrium conditions until the velocity of the more dense rods structure suppress it. The nonequilibrium crystallization becomes possible due to the expansion of the region of existence of BCC-structure in presence of moving boundaries and respective kinetic-liquidus offset. We encountered a number of difficulties in the front determination for the low driving forces, so the densities correspondent to the solidus line was evaluated to achieve the stable BCC-crystal. 
The sequence of growth speeds of BCC-crystal preserves for most initial parameters. The form of the curves is the same as it follows from the general $tanh$-solution of PFC-amplitude equations~\cite{gse,niz18}. With the increase of driving force the velocity tends to its asymptotic limit given by $V\to V_\phi$ with $V_\phi \sim 1/\sqrt{\tau}$. From Fig.~\ref{vel} follows that the main difference in the predictions of the parabolic PFC-model ($\tau=0$) and the hyperbolic PFC-model ($\tau>0$) consists in the decreasing of the front velocity with the increasing $\tau$. This occurs due to additional degree of freedom, i.e. the atomic flux as independent variable in the hyperbolic model, which need to additionally relax in comparison with the parabolic ($\tau=0$) model.


\begin{figure}
  \includegraphics[width=\columnwidth]{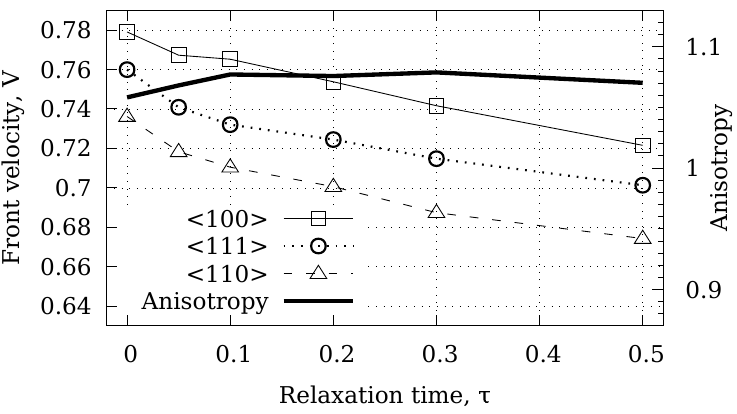}
  \caption{\label{vtau} BCC front velocities $V$ as a function of relaxation time $\tau$ for different crystallographic faces. Represented lines and points correspond to initial density $n_0=-0.36$ and driving force $\varepsilon=0.4$. The growth velocity anisotropy $V_{\langle 100 \rangle}/V_{\langle 110 \rangle}$ is shown by the bold line; values presented on the right y-axis.  }
  \end{figure}

The dependence of front velocity on the value of relaxation time presented in Fig.~\ref{vtau} shows the nonlinear decreasing of $V(\tau)$. This figure also exhibits a small effect of anisotropies on $\tau$ for a fixed $\varepsilon$. Apparently, the contribution of flux term in Eq.~(\ref{dyn}) is isotropic and influences on each crystal face identically.

Obtained growth anisotropies and thus difference in velocities for low initial density $n_0$ shown in Fig.~\ref{frontPos} begin to decrease as the initial density $n_0$ rises. Increasing of $n_0$ makes anisotropy smaller [$\approx 1$, see point $\varepsilon=0.4$, for $n_0=-0.31$ of Fig.~\ref{vel}, where $V_{\langle 100 \rangle}=1.4204$, $V_{\langle 111 \rangle}=1.42$, $V_{\langle 110 \rangle}=1.41$] due to effective increasing of the gradient of chemical potential. The anisotropy of growth velocities changed from 1 up to 1.06   for intermediate $\varepsilon$.

\subsection{Reticular density}~\label{retdens}
The growth morphology and movement of crystal lattice faces are determined by the  lattice type. According to the Bravais empirical law the order of preferable growth directions for faces depends on the reticular density~\cite{Sunagawa2007}: the faster growth exists for the faces with the lower reticular densits of the specific face. The reticular density $\rho_{\langle \texttt{hkl} \rangle}$ is defined as the number of atoms (or its fraction) per unit area on a plane~\cite{Chadwick1967}.

Let us summarize the results for reticular density of two dimensional triangle structure and three dimensional face centered cubic (FCC) and BCC lattices using the geometrical approach. For the equilibrium lattice parameter $a$
the correspondent densities would be (from lowest to highest value of the density)
  \begin{align*}
  \textrm{1) Triangle:} \quad \!\!
  \rho_{\langle 10 \rangle}=\frac{1}{3 \sqrt{3} a}; \quad
  \rho_{\langle 11 \rangle}=  \frac{1}{2 a};
\end{align*}
 \begin{equation*}
 \textrm{2) for FCC:} \quad \!\!
  \rho_{\langle 110 \rangle}=\frac{\sqrt{2}}{a^2}; \quad
  \rho_{\langle 100 \rangle}=\frac{2}{a^2}; \quad
  \rho_{\langle 111 \rangle}=\frac{4}{\sqrt{3}a^2};
\end{equation*}
 \begin{equation*}
  \textrm{3) for BCC:} \quad \!\!
  \rho_{\langle 100 \rangle}=\frac{1}{a^2}; \quad
  \rho_{\langle 111 \rangle}=\frac{19 \sqrt{3}}{27 a^2}; \quad
  \rho_{\langle 110 \rangle}=\frac{\sqrt{2}}{a^2}.
\end{equation*}
Therefore, according to the Bravais law, the BCC structure should exhibit the following sequence for the growth velocity $V_{\langle 100 \rangle}>V_{\langle 111 \rangle}>V_{\langle 110 \rangle}$ that agrees well with the calculations of the present work (see results of Fig.~\ref{frontPos} and Fig.~\ref{vel}).

\subsection{Formation of disordered crystals}
\begin{figure}
  \includegraphics[width=\columnwidth]{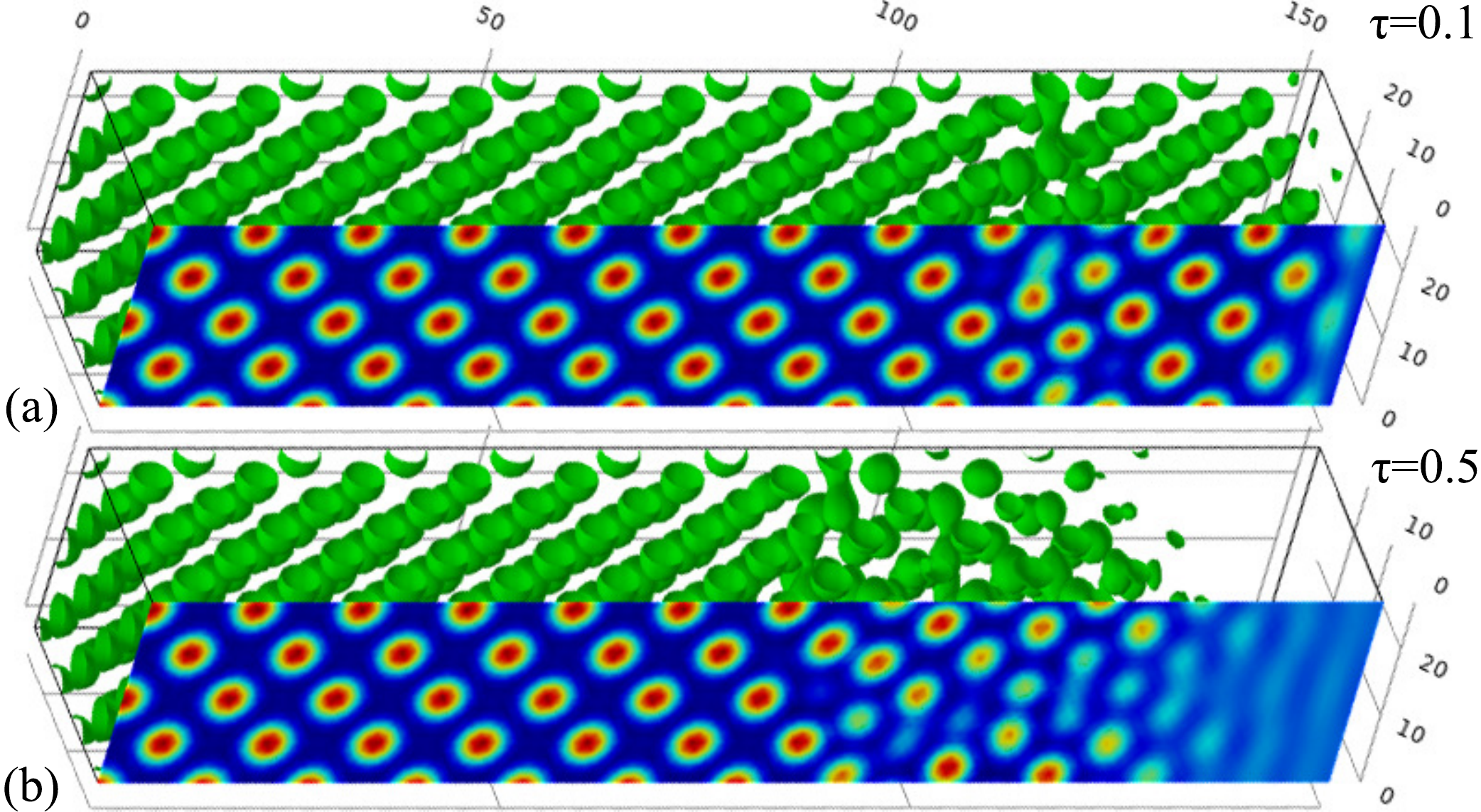}
  \caption{\label{defect} Snapshots  of propagating fronts (for different $\tau$) showing the relaxation of defect-structures selected on rapid front along the $\langle 110 \rangle$  at $t=100$ at high driving forces $\varepsilon=0.5$, $n_0=-0.38$.} \end{figure}

Originating of disordered crystal structures is a special task of experimental and theoretical works (see Ref.~\cite{garosoc} and references therein). In the present study, large values of $\varepsilon$ and $\tau$ could lead to the formation of a disordered structure due to capturing of atomic disorder from the liquid, see Fig.~\ref{defect}. With this process the specific face decelerates and becomes wider. This defect structure was also previously observed, in two-dimensional case of solidification from an supercooled liquid~\cite{Archer2012,Granasy2019}. This effect is explained as a consequence of the mismatch between the selected lattice parameter  and the equilibrium lattice spacing. The mismatch increases with increasing of driving force and in case of present hyperbolic PFC-equation, with the increasing of the relaxation time $\tau$. The marginal stability analysis could explain these defects (dislocations), emerging presumably due to the stress arising from the non-equilibrium lattice constant selected at the rapidly growth crystal face~\cite{gse,ge-2011}.
Computation and quantitative estimations of disordered zone around solidification front are of special interest of analysis \cite{Galenko2018}. A search for quantitative estimations of this disorder in the PFC-model is a perspective topic of future study.

\section{Conclusions}

The hyperbolic (modified) PFC-model which takes into account relaxation of the phase field and relaxation of the flux of atomic density has been used in numerical study of rapidly growing $\langle 100 \rangle$, $\langle 110 \rangle$ and $\langle 111 \rangle$ faces in body centered cubic crystal lattice (BCC lattice). In particular, an influence of the local relaxation time and the effect of atomic reticular density of different crystal faces on the interface velocity has been studied.

The PFC-modeling shows that the BCC lattice exhibits the following sequence for the growth velocity $V_{\langle 100 \rangle}>V_{\langle 111 \rangle}>V_{\langle 110 \rangle}$ that agrees well with the experimental Bravais law accordingly which the faster growth exists for the faces with the lower reticular densities of the specific face. Formation of the disordered structure due to the lack of time for the local relaxation of structure has been obtained in modeling of different faces at the growth under high driving forces and relaxation times.

\section*{Acknowledgments}
The support by Russian Science Foundation (Funder Id http://dx.doi.org/10.13039/501100006769) \, under \, Grant RSF~18-12-00438 is acknowledged.


\bibliography{arxiv}

\end{document}